\documentclass[a4paper,11pt]{article}
\pdfoutput=1 

\usepackage{jheppub} 

\usepackage[T1]{fontenc} 

\title{\boldmath Local thermal behaviour of a massive scalar field near a reflecting wall}

\author[a]{V. A. De Lorenci,}
\author[b]{L. G. Gomes}
\author[b,1]{and E. S. Moreira Jr.\note{Corresponding author.}}

\affiliation[a]{Instituto de F\'{\i}sica e Qu\'{i}mica, 
Universidade Federal de Itajub\'a, 
Itajub\'a, MG 37500-903, Brazil}
\affiliation[b]{Instituto de Matem\'{a}tica e Computa\c{c}\~{a}o, 
Universidade Federal de Itajub\'a, 
Itajub\'a, MG 37500-903, Brazil}

\emailAdd{delorenci@unifei.edu.br}
\emailAdd{lggomes@unifei.edu.br}
\emailAdd{moreira@unifei.edu.br}

\abstract{ The mean square fluctuation and the expectation value of the stress-energy-momentum tensor
of a neutral massive scalar field at finite temperature are determined near an infinite plane Dirichlet wall, and also near an infinite plane Neumann wall. The flat background has an arbitrary number of dimensions and the field is arbitrarily coupled to 
the vanishing curvature. It is shown that, unlike vacuum contributions, 
thermal contributions are free from boundary divergences,
and  that the thermal behaviour of the scalar field 
near a  Dirichlet wall differs considerably from that near a Neumann wall. 
Far from the wall the study reveals
a local version of dimensional reduction, namely,  
corrections to  familiar blackbody expressions 
are linear in the temperature, with the corresponding coefficients given only in terms of vacuum expectation values in a background with one less dimension. 
It is shown that such corrections are ``classical'' (i.e., not dependent on  Planck's constant) only if the scalar field is massless.
A natural conjecture that arises is that
the ``local dimensional reduction'' is universal
since it operates for massless and massive fields alike and regardless of the boundary conditions.

}


\arxivnumber{1410.7826}

\begin{document} 
\maketitle
\flushbottom

\section{Introduction}
\label{introduction}
After the recent discovery of the Higgs boson, 
interest on the physics of massive scalar fields became rather more phenomenological and less academic. As is well known, there are 
models involving scalar fields in cosmology and high energy physics
where a natural protagonist is the Higgs boson (see e.g. ref. \cite{bez08}). If one adds that primordial hot radiation permeates the universe  and
that modern colliders  recreates  the extremely hot conditions just after the big bang, in this scenario it seems pertinent to study the  thermal properties of a massive scalar field
in great detail.

The tools to study scalar radiation at temperature $T$ in an infinite 
cavity do not differ from those used in dealing with the familiar electromagnetic blackbody radiation. One can for example identify the scalar radiation in the cavity with an ideal gas of bosons  with zero chemical potential, leading to the well known 
distribution
\begin{equation}
n_{\bf p}=
\frac{1}{\exp{\left\{\beta\sqrt{(pc)^2+(Mc^2)^2}\right\}-1}},
\label{n}
\end{equation}  
where $\beta:=1/k_{B}T$ as usual.
Setting $M=0$ and considering integrations over phase space and momentum, eq. (\ref{n}) yields  the familiar Planckian expressions for the energy density $\rho$ 
and  pressure $P$, 
\begin{eqnarray}
&&
\rho=\frac{\pi^2}{30}\frac{(k_{B}T)^4}{(\hbar c)^3},
\hspace{1cm}
P=\frac{\rho}{3},
\label{gas}
\end{eqnarray}
corresponding to the following uniform and isotropic
stress-energy-momentum tensor
\begin{equation}
\left<T^\mu{}^\nu\right> ={\rm diag}(\rho,P,P,P).
\label{semt}
\end{equation}

One might also consider the situation that, although still dealing with  an infinite cavity, the interest is in the scalar radiation near, say, a plane wall of the cavity (with the other walls at infinity).
Applying local techniques of field theory at finite temperature $T$, 
assuming that the scalar field is massless, conformally coupled
(i.e., with curvature coupling parameter $\xi=1/6$) and that it satisfies the Dirichlet boundary condition on the wall, Kennedy, Critchley and Dowker \cite{ken80} have found the following
ensemble average near the infinite wall lying on a Cartesian plane, 
\begin{equation}
\left<T^\mu{}^\nu\right> ={\rm diag}\left(\frac{\rho}{9},P,-\frac{P}{3},-\frac{P}{3}\right),
\label{dowker}
\end{equation}
where $\rho$ and $P$ are those in eq. (\ref{gas}) and subleading contributions (that vanish when the distance to the plane wall vanishes) have been neglected. In fact, only the second component in eq. (\ref{dowker}) --- pressure on the wall --- is strictly uniform
as have been shown by Tadaki and Takagi \cite{tad86}.

The source of apparent conflict between eqs. (\ref{semt}) and (\ref{dowker})
is that by assuming the distribution (\ref{n}) one is ignoring the presence of walls. Boundaries ``deform'' the vacuum
(Casimir-like effects \cite{mil01,bor01}) 
and modify  ensemble averages. It follows that
$\left<T^\mu{}^\nu\right>$ in  eq. (\ref{semt}) is correct only if one is considering scalar radiation deep in the bulk, i.e., far away from the walls of the cavity
(or, equivalently, at high temperature) \cite{ken80,tad86}.

By investigating hot radiation near curved 
boundaries, Balian and Duplantier \cite{bal78} have shown that
similar kind of non trivial local effects also arise with the electromagnetic radiation. In particular they have
shown that the energy density presents a non integrable divergence
when a curved wall of a perfect conductor is approached.
Boundary divergences were also investigated by 
DeWitt in the classic text \cite{dew75} concluding
that even near a reflecting plane wall
the vacuum expectation value $\left<T^\mu{}^\nu\right>$ of a 
non conformally coupled scalar field diverges at the wall. 
In fact, for an arbitrary $\xi$
eq. (\ref{dowker}) is replaced by \cite{tad86}
\begin{equation}
\left<T^\mu{}^\nu\right> ={\rm diag}
\left(\rho_0+\frac{(1-4\xi)\rho}{3},P,
-\rho_0-(1-4\xi)P,-\rho_0-(1-4\xi)P\right),
\label{de}
\end{equation}
where the vacuum energy density
\begin{equation}
\rho_0:=-\frac{(1-6\xi)\hbar c}{16\pi^2x^4}
\label{vedensity}
\end{equation}
carries  the non integrable divergence arising when the
Dirichlet wall is approached ($x\rightarrow 0$).

Boundary divergences were interpreted by Deutsch and Candelas \cite{deu79} 
as a consequence of oversimplification  of real boundaries, and that has
at some extend  been confirmed along the years by further investigations \cite{bar84,dav85,gra04,ful10,mil11,bou12,maz11,bar12}. 
To keep these divergences under control one can use certain ``cut-off procedures'' according to which vacuum expectation values such as $\rho_0$ in eq. (\ref{vedensity})
hold close, but not that close to the wall.
An alternative and interesting way of dealing with  boundary divergences is the ``renormalization procedure'' suggested in ref. \cite{ken80} (see also ref. \cite{mil14}).
Within this approach $\rho_0$ in eq. (\ref{vedensity})
hides a $\delta$-function contribution which, 
after integration over space, gives rise to a
surface energy that kills off the divergent contribution in the formal expression for the total vacuum energy, resulting that the latter becomes finite. The ``renormalization procedure'' was 
extended to the electromagnetic field in ref. \cite{ken82}, and to deal with
more general boundary conditions for the massless scalar field on plane boundaries in ref. \cite{rom02}.
Also worth mentioning is a ``cut-off procedure'' (possibly related with the ``renormalization procedure'' just considered) that has been proposed by Ford and Svaiter \cite{for98}. It consists of taking into account the quantum nature of the boundary, resulting that the ``width'' of the boundary's wave packet becomes a regulator parameter.

Investigations on the global thermal behaviour of the scalar field in backgrounds with boundaries go back to the 1970's \cite{dow78,amb83} and since then the subject has been much considered in the literature 
(see reviews \cite{mil01,bor01} and refs. \cite{kir91a,kir91b,fei01,sca06,lim07,gey08,web09,teo09}), unlike its local thermal behaviour that  has been much less studied. 
Early investigations on the local thermal behaviour of a scalar field near a reflecting wall 
have been restricted to  massless fields 
\cite{ken80,tad86}.
The present paper extends the study to examine a massive field.
In so doing, the philosophy in the work by Brown and Maclay
\cite{bro69} to deal with the electromagnetic field between conducting plates is followed. Namely, the ensemble average of 
a quantity $\cal{A}$ will be written as,
\begin{equation}
\left<\cal{A}\right>=
\left<\cal{A}\right>_{{\tt vacuum}}
+
\left<\cal{A}\right>_{{\tt mixed}}
+
\left<\cal{A}\right>_{{\tt thermal}},
\label{bm}
\end{equation} 
where 
$\left<\cal{A}\right>_{{\tt vacuum}}$
is the vacuum expectation value of $\cal{A}$ 
(obtained by setting the temperature equal to zero in eq. (\ref{bm})),
$\left<\cal{A}\right>_{{\tt thermal}}$ is
the blackbody contribution (that obtained by using eq. (\ref{n})), and 
$\left<\cal{A}\right>_{{\tt mixed}}$ is 
a ``mixed'' contribution (with vacuum-thermal nature) that connects the two others such that it vanishes at zero temperature and far away from the reflecting wall.

To arrive to $\left<\cal{A}\right>$ for a massive and arbitrarily coupled scalar field $\phi$, the ``point-splitting'' approach 
\cite{dav82,ful89} is applied to the Schwinger ``proper time'' representation of the Feynman propagator (see e.g. ref. \cite{deu79}) at finite temperature. The latter is calculated in appendix  \ref{ftpropagator} for a flat $N$-dimensional spacetime containing an infinite plane wall where
$\phi$ satisfies the Dirichlet (or Neumann) boundary condition. 
In section \ref{rpropagator}, the Feynman propagator is
renormalized giving rise to an expression involving infinite sums of modified Bessel functions of the second kind $K_{\nu}({\rm z})$, where $\nu$  is fixed by the dimensionality $N$ of the spacetime. 
The ensemble averages $\left<\phi^2\right>$ and $\left<T^\mu{}^\nu\right>$ are found and investigated
in sections \ref{fluctuation} and \ref{stress}, respectively.
In section \ref{high temperature}, the leading order  behaviours of $\left<\phi^2\right>$ and $\left<T^\mu{}^\nu\right>$ for an ``ultralight'' scalar field (i.e., the corresponding massless expressions)  are briefly addressed.
(It should be pointed out that $\left<T^\mu{}^\nu\right>$ for  hot  massless scalar radiation near a reflecting wall has been studied previously \cite{ken80,tad86,lor14}; whereas no record was found on  $\left<\phi^2\right>$.) 
In section \ref{neumann}, the arguments developed in the previous 
sections for the Dirichlet boundary condition are extended to 
consider Neumann's boundary condition, and 
a parallel between the corresponding thermal behaviours is drawn.
Section \ref{conclusion} contains a summary and some remarks on the ensemble averages found in the text, ending with
proposals on further study of 
a local version of the dimensional reduction,  
noted in this work, which
relates vacuum averages in $N-1$ dimensions to
(in general ``non-classical'')  corrections 
to blackbody expressions in $N$ dimensions. Appendix \ref{aexpressions} contains alternative expressions 
that turned out to be useful in the text.  
(Unless stated otherwise, dimensions are such that $k_{B}=\hbar=c=1$.)

\section{Renormalized propagator}
\label{rpropagator}

Consider an infinite cavity in an $N$-dimensional flat spacetime
(cf. eq. (\ref{metric1})). One of the walls of the $(N-1)$-dimensional cavity coincides with the plane $x=0$, and the other walls are at infinity. A point in the 
spacetime is labelled by flat coordinates
$(t,x,y,z,\cdots)$ where $|x|$ is the distance to the plane wall. Due to the obvious symmetry of the background, $x$ is taken to be non negative in the rest of the text. A neutral scalar field
$\phi$ with mass $M$ is assumed to be confined to the cavity and in thermodynamic equilibrium with the walls at temperature $T$. To prevent fluxes through the plane wall, a typical boundary condition used is Dirichlet's which will be taken here. Thus $\phi=0$ at  $x=0$.
(Neumann's boundary condition is addressed in sections \ref{neumann} and \ref{conclusion}.)

In order to obtain the  Feynman propagator 
$G_{{\cal F}}({\rm x},{\rm x}')$ at finite temperature $T$ one solves
\begin{equation}
\left(\Box_{{\rm x}}+M^{2}\right)G_{{\cal F}}({\rm x},{\rm x}')=
-\delta\left({\rm x}-{\rm x}'\right),
\label{fe0}
\end{equation}
observing the usual prescription (see e.g. ref. \cite{ful87})
of continuing time $t$ to imaginary values, taking $it$
periodic with period $1/T$. This is done in appendix \ref{ftpropagator}, and the result is
\begin{eqnarray}
&&G_{{\cal F}}({\rm x},{\rm x}')=
-\frac{i}{(2\pi)^{N/2}}M^{\frac{N-2}{2}}
\nonumber
\\
&&\times
\sum_{n=-\infty}^{\infty}
\left[(-\sigma_{-})^{\frac{2-N}{4}}
K_{\frac{N-2}{2}}\left(M\sqrt{-\sigma_{-}}\right)
-
(-\sigma_{+})^{\frac{2-N}{4}}
K_{\frac{N-2}{2}}\left(M\sqrt{-\sigma_{+}}\right)
\right],
\label{pro}
\end{eqnarray}
where 
$\sigma_{\pm}:=(t-t'-in/T)^2-(x\pm x')^2-(y-y')^2-(z-z')^2-\cdots$.
Recalling that for small argument and $\nu>0$,
\begin{equation}
K_{\nu}({\rm z})=2^{\nu-1}\Gamma(\nu)
{\rm z}^{-\nu}+\cdots,
\label{sarguments}
\end{equation}
by setting $M\rightarrow 0$ 
in eq. (\ref{pro}) the massless propagator is recovered  \cite{lor14}. 

By studying eq. (\ref{pro}), one sees that the term corresponding to $n=0$ and involving $\sigma_{-}$ is simply the familiar vacuum propagator in Minkowski spacetime $G_{0}({\rm x},{\rm x}')$. That corresponding
to $n=0$ but now involving  $\sigma_{+}$ is the vacuum propagator in the
presence of the Dirichlet wall $G_{{\tt vacuum}}({\rm x},{\rm x}')$. 
The first sum $\sum_{n\neq 0}$ in eq. (\ref{pro}) yields the thermal propagator in Minkowski spacetime  $G_{{\tt thermal}}({\rm x},{\rm x}')$,
while the second sum $\sum_{n\neq 0}$ yields a propagator $G_{{\tt mixed}}({\rm x},{\rm x}')$ of a mixed (vacuum-thermal) nature. Therefore eq. (\ref{pro}) can be cast as
\begin{equation}
G_{{\cal F}}({\rm x},{\rm x}')=G_{0}({\rm x},{\rm x}')
+G_{{\tt vacuum}}({\rm x},{\rm x}')
+G_{{\tt mixed}}({\rm x},{\rm x}')
+G_{{\tt thermal}}({\rm x},{\rm x}').
\label{fp}
\end{equation}
Since spacetime is flat one renormalizes by removing the zero temperature Minkowski contribution, resulting the renormalized propagator, 
\begin{equation}
G({\rm x},{\rm x}')=
G_{{\tt vacuum}}({\rm x},{\rm x}')
+G_{{\tt mixed}}({\rm x},{\rm x}')
+G_{{\tt thermal}}({\rm x},{\rm x}').
\label{rfp}
\end{equation}
As averages can formally be obtained by linear operations in the renormalized propagator \cite{dav82,ful89}, the expression for
the ensemble average  $\left<\cal{A}\right>$ in  eq. (\ref{bm}) follows from eq. (\ref{rfp}). Taking into account that for large argument $K_{\nu}({\rm z})$ falls exponentially,
the following limits can be readily verified,
\begin{equation}
\lim_{T\rightarrow 0}G({\rm x},{\rm x}')=G_{{\tt vacuum}}({\rm x},{\rm x}'),
\hspace{2.0cm}
\lim_{x,x'\rightarrow \infty}G({\rm x},{\rm x}')=G_{{\tt thermal}}({\rm x},{\rm x}'). 
\label{limits}
\end{equation}
In both these limits 
$G_{{\tt mixed}}({\rm x},{\rm x}')$ vanishes, and all that still holds if the field is massless with $N>3$. (For $N\leq 3$
when $M=0$, divergences typical of lower dimensions appear \cite{ful87}. More on that in section \ref{high temperature}.)

\section{Mean square field fluctuation}
\label{fluctuation}
The ensemble average 
$\left<\phi^2\right>$ measures how much the scalar field $\phi$ fluctuates around $\left<\phi\right>=0$. It is obtained from eq. (\ref{rfp}) simply by considering 
$\left<\phi^2\right>=iG({\rm x},{\rm x})$, 
\begin{equation}
\langle\phi ^{2}\rangle^{(N)}=
\langle\phi^{2}\rangle_{{\tt vacuum}}^{(N)}
+\langle\phi ^{2}\rangle_{{\tt mixed}}^{(N)}+
\langle\phi ^{2}\rangle_{{\tt thermal}}^{(N)},
\label{phi2}
\end{equation}
where the superscript $(N)$ is not an exponent but simply indicates dimensionality, which will play special role below. 
The last term in eq. (\ref{phi2}) is the uniform blackbody contribution,
\begin{equation}
\langle\phi ^{2}\rangle_{{\tt thermal}}^{(N)}=
\frac{1}{\pi}\left(\frac{MT}{2\pi}\right)^{\frac{N-2}{2}}
\sum_{n=1}^{\infty}n^{\frac{2-N}{2}}K_{\frac{N-2}{2}}
\left(\frac{Mn}{T}\right).
\label{phi2t}
\end{equation}
(For a quick check of eq. (\ref{phi2t})
one sets $N=4$ and $M\rightarrow 0$
observing eq. (\ref{sarguments}), then the sum yields $\zeta(2)$ leading to the familiar expression $T^{2}/12$.)
The first term is the vacuum fluctuation,
\begin{equation}
\langle\phi ^{2}\rangle_{{\tt vacuum}}^{(N)}=
-\frac{1}{2^{N-1}\pi^{N/2}}\left(\frac{M}{x}\right)^{\frac{N-2}{2}}
K_{\frac{N-2}{2}}
\left(2Mx\right),
\label{phi2v}
\end{equation}
where by setting $N=4$  and $M\rightarrow 0$ one recovers the well known
expression $-1/16\pi^2 x^2$ \cite{ful89}. Equation (\ref{phi2v}) also reproduces the result in ref. \cite{hil86} where the Schr\"{o}dinger formalism to obtain vacuum averages for massive scalar fields near a reflecting wall has been used.

The mixed fluctuation in eq. (\ref{phi2}) is given by
\begin{equation}
\langle\phi ^{2}\rangle_{{\tt mixed}}^{(N)}=
-\frac{1}{\pi}\left(\frac{MT}{2\pi}\right)^{\frac{N-2}{2}}
\sum_{n=1}^{\infty}
\left[(2Tx)^2+n^2\right]^{\frac{2-N}{4}}K_{\frac{N-2}{2}}
\left(\frac{M}{T}\sqrt{(2Tx)^2+n^2}\right).
\label{phi2m}
\end{equation}
It is a kind of non uniform ``thermal'' contribution that satisfies 
\begin{eqnarray}
\langle\phi ^{2}\rangle_{{\tt mixed}}^{(N)}=
-\langle\phi ^{2}\rangle_{{\tt thermal}}^{(N)},
&\qquad& x=0.
\label{lowt}
\end{eqnarray}

Since $\left<\phi^2\right>$ and $\left<T^\mu{}^\nu\right>$
share many common features,
it is pedagogical to study $\left<\phi^2\right>$ more closely.
Unlike the vacuum fluctuation in eq. (\ref{phi2v}) that diverges
as $x^{2-N}$ (see eq. (\ref{sarguments})) when the wall is approached, the mixed contribution is free from boundary divergences (cf. eq. (\ref{lowt})).
(Note that the thermal contribution is also divergence free since it is uniform.) As mentioned earlier, if one is interested in obtaining global quantities by integranting local quantities over space, then ``cut-off procedures'' are required to regulate integrations. It should be pointed out however that, whichever the ``cut-off procedure'' used, expressions for vacuum expectation values, such as that in eq. (\ref{phi2v}), are reliable as long as one is not too close to the reflecting wall.

By moving deep in the bulk of the cavity (i.e., $x\rightarrow\infty$), clearly the only contribution left behind in eq. (\ref{phi2}) is the blackbody contribution 
$\langle\phi ^{2}\rangle_{{\tt thermal}}^{(N)}$.
On the other hand, by progressively approaching the wall ($x\rightarrow 0$) only the (divergent)
$\langle\phi ^{2}\rangle_{{\tt vacuum}}^{(N)}$
is left at the end (see eqs. ({\ref{phi2}) and (\ref{lowt})).
It should be stressed that the effects resulting of moving deep in the bulk
and near the wall
are equivalent, respectively,  to those obtaining by keeping $x$ fixed raising the temperature ($T \rightarrow \infty$) and lowering it
($T \rightarrow 0$).
Note that all these features are consistent with the asymptotic behaviour of the renormalized propagator (see eq. (\ref{limits}) and text just following it).

In order to find the leading correction to 
$\langle\phi^{2}\rangle_{{\tt vacuum}}^{(N)}$ when $Tx\ll 1$
(low temperature or near the wall), the expression for 
$\langle\phi ^{2}\rangle_{{\tt mixed}}^{(N)}$ in eq. (\ref{phi2m})
is expanded in powers of $Tx$ whose leading term is that in
eq. (\ref{lowt}). By considering derivatives of $K_{\nu}({\rm z})$, identities relating these functions \cite{gra07}, and
omitting terms of higher powers, eq. (\ref{phi2}) leads to 
\begin{eqnarray}
\langle\phi ^{2}\rangle^{(N)}=
\langle\phi^{2}\rangle_{{\tt vacuum}}^{(N)}+
4\pi x^{2}\langle\phi ^{2}\rangle_{{\tt thermal}}^{(N+2)},
&\qquad &
Tx\ll 1.
\label{lowtphi2}
\end{eqnarray}
Thus, at low temperature or near the wall, the leading correction to the vacuum fluctuation in $N$ dimensions is determined from the thermal fluctuation in $N+2$ dimensions (see eq. (\ref{phi2t})).


The leading correction to 
$\langle\phi^{2}\rangle_{{\tt thermal}}^{(N)}$ when $Tx\gg 1$
(high temperature or deep in the bulk)
can be obtained by replacing the sum in eq. (\ref{phi2m}) by an integration. The latter can be evaluated \cite{gra07}, 
resulting in
\begin{eqnarray}
\langle\phi ^{2}\rangle^{(N)}=
\langle\phi ^{2}\rangle^{(N)}_{{\tt thermal}}+
\langle\phi ^{2}\rangle^{(N)}_{{\tt class}},&\qquad &
Tx\gg 1,
\label{hol1}
\end{eqnarray}
where after reintroducing dimensionful $\hbar$ and $c$,
\begin{equation}
\langle\phi ^{2}\rangle^{(N)}_{{\tt class}}:= 
\langle\phi ^{2}\rangle^{(N-1)}_{{\tt vacuum}}\ \frac{T}{\hbar c}.
\label{classvf}
\end{equation}
It should be pointed out that eq. (\ref{hol1})
is exact up to exponentially small corrections 
which vanish as $Tx\rightarrow \infty$
(note that $\langle\phi ^{2}\rangle^{(N)}_{{\tt vacuum}}$
in eq. (\ref{phi2}) is cancelled by the subleading contribution in eq. (\ref{phi2m})),
and that it holds also for $N=2$ by considering $N=1$ 
in eq. (\ref{phi2v}). The behaviour corresponding to eqs. (\ref{hol1}) and (\ref{classvf}) is the first example in this paper of a local version of dimensional reduction. As will be seen in the next section, such a dimensional reduction will also be present in the behaviour of $\left<T^\mu{}^\nu\right>$.

As the dimensionful factor $1/\hbar c$ appears in eq. (\ref{classvf}), accordingly eq. (\ref{phi2v})
must be multiplied by $\hbar^{2-N/2}c^{N/2}$ and the argument of the corresponding Bessel function 
in eq. (\ref{phi2v}) must be 
multiplied by $c/\hbar$. Taking into account these amendments, 
it is  easy to verify from eq. (\ref{classvf}) that in general the correction 
$\langle\phi ^{2}\rangle^{(N)}_{{\tt class}}$ depends on $\hbar$, and will be  ``classical'' only if the field is massless: if $M\rightarrow 0$, because of eq. (\ref{sarguments}), massless 
$\langle\phi ^{2}\rangle^{(N)}_{{\tt vacuum}}$ 
will depend on $\hbar$ and $c$ only through an overall factor
$\hbar c$ which will cancel that in eq. (\ref{classvf}). To illustrate this fact one sets $N=4$ in eq. (\ref{classvf}) noting that 
$K_{1/2}({\rm z})=\sqrt{(\pi/2{\rm z})}\exp(-{\rm z})$. It follows then 
\begin{equation}
\langle\phi ^{2}\rangle^{(N=4)}_{{\tt class}}= 
-\frac{T}{8\pi x}\exp\left(-\frac{2Mcx}{\hbar}\right),
\label{4classvf}
\end{equation}
which will clearly be classical only if $M=0$.
At this point it should be remarked that in the study of the high temperature behaviour of massless fields, ``classical'' corrections linear in the temperature to blackbody expressions are a long known feature \cite{bro69,bal78,fei01,sca06}. The result above suggests that
the designation ``classical'' for such corrections is not appropriate when the field has non-vanishing mass.

\section{Stress-energy-momentum tensor}
\label{stress}
The local content of energy and momentum, as well as the stresses, 
of the hot scalar radiation in the infinity cavity
are given by the ensemble average 
$\left<T^\mu{}^\nu\right>$, which can be formally obtained by acting with the differential operator
\begin{equation}
{\cal D}^\mu{}^\nu({\rm x},{\rm x}'):=
(1-2\xi)\partial^\mu\partial^{\nu'}- 2\xi\partial^\mu\partial^{\nu} 
+ (2\xi - 1/2)\eta^\mu{}^\nu
\partial^\lambda\partial_{\lambda'} 
+2M^2\eta^\mu{}^\nu(1/4-\xi)
\label{doperator}
\end{equation}
on the renormalized propagator in eq. (\ref{rfp}),
\begin{eqnarray}
\left<T^\mu{}^\nu\right> &&= i \lim_{{\rm x}'\rightarrow {\rm x}}
{\cal D}^\mu{}^\nu({\rm x},{\rm x}')\ G({\rm x},{\rm x}')
\nonumber
\\
&&=
\left<T^\mu{}^\nu\right>_{{\tt vacuum}}+
\left<T^\mu{}^\nu\right>_{{\tt mixed}}+
\left<T^\mu{}^\nu\right>_{{\tt thermal}},
\label{prescription}
\end{eqnarray}
resulting,
\begin{equation}
\left<T^\mu{}^\nu\right>^{(N)} ={\rm diag}(\rho^{(N)},P_{\perp}^{(N)},P_{\parallel}^{(N)},\cdots,P_{\parallel}^{(N)}),
\label{emt}
\end{equation} 
where the terms in eq. (\ref{prescription}) are all diagonal 
and each one of the elements in eq. (\ref{emt}) has the form in eq. (\ref{bm}).
Accordingly, the energy density $\rho^{(N)}$ is given by
\begin{equation}
\rho^{(N)}=
\rho^{(N)}_{{\tt vacuum}}
+
\rho^{(N)}_{{\tt mixed}}
+
\rho^{(N)}_{{\tt thermal}},
\label{energy}
\end{equation} 
with the blackbody energy density $\rho^{(N)}_{{\tt thermal}}$
and the isotropic blackbody radiation pressure,
\begin{equation}
P_{\perp\ {\tt thermal}}^{(N)}\equiv
P_{\parallel\ {\tt thermal}}^{(N)}
= 2\pi \langle\phi ^{2}\rangle_{{\tt thermal}}^{(N+2)},
\label{rpressure}
\end{equation}
related by the equation of state
\begin{equation}
\rho^{(N)}_{{\tt thermal}}=(N-1)P_{\perp\ {\tt thermal}}^{(N)}
+M^2\langle\phi ^{2}\rangle_{{\tt thermal}}^{(N)}.
\label{eqstate}
\end{equation}
Observing eq. (\ref{sarguments}), by setting $N=4$ and $M\rightarrow 0$ in eqs. (\ref{rpressure}) and (\ref{eqstate}) one reproduces eq. (\ref{gas}) as it should. Note that eq. (\ref{rpressure}) 
relates  blackbody radiation pressure in $N$ dimensions with thermal fluctuation in $N+2$ dimensions (see eq. (\ref{phi2t})).

The vacuum energy density in eq. (\ref{energy}) 
can be written in terms of vacuum fluctuations in eq. (\ref{phi2v}),
\begin{equation}
\rho^{(N)}_{{\tt vacuum}}=8\pi(1-N)(\xi-\xi_{N})
\langle\phi ^{2}\rangle_{{\tt vacuum}}^{(N+2)}
+M^2(1-4\xi)\langle\phi ^{2}\rangle_{{\tt vacuum}}^{(N)},
\label{venergy}
\end{equation}
with ($\xi=\xi_{N}$ is the conformal coupling)
\begin{equation}
\xi_{N}:=\frac{N-2}{4(N-1)}, 
\label{cc}
\end{equation}
and it satisfies the following equation of state,
\begin{equation}
P_{\parallel\ {\tt vacuum}}^{(N)}=-\rho^{(N)}_{{\tt vacuum}},
\label{veqstate}
\end{equation}
agreeing again with early calculations \cite{hil86,mel11}
(it should be reported that an overall $-1$ factor is missing in eq. (B.11) of ref. \cite{hil86}).
Note that, at the wall, the vacuum energy density diverges as   
$x^{-N}$ if $\xi\neq\xi_{N}$, and as  
$x^{2-N}$ if $\xi=\xi_{N}$ with $M\neq 0$.

To complete the expression for the energy density in 
eq. (\ref{energy}), the mixed contribution is
(see also eq. (\ref{aexpression1}))
\begin{eqnarray}
\rho^{(N)}_{{\tt mixed}}=&&
2\left(\frac{MT}{2\pi}\right)^{N/2}
\sum_{n=1}^{\infty}\left[(2Tx)^2+n^2\right]^{-(N+4)/4}
\hspace{6.5cm}
\nonumber
\\
&&\times
\left\{4
\left[(N-1)(\xi-\xi_N)\left(4T^2x^2+n^2\right)-\xi Nn^2\right]
K_{\frac{N}{2}}
\left(\frac{M}{T}\sqrt{(2Tx)^2+n^2}\right)\right. 
\hspace{1.4cm}
\nonumber
\\
&&
\hspace{0.1cm}
\left. +
\frac{M}{T}\left[(4Tx)^2(\xi-1/4)-n^2\right]
\sqrt{(2Tx)^2+n^2}
K_{\frac{N-2}{2}}
\left(\frac{M}{T}\sqrt{(2Tx)^2+n^2}
\right)\right\}.
\label{mixed00}
\end{eqnarray}
As happens to all the  mixed contributions in this paper,
$\rho^{(N)}_{{\tt mixed}}$ is finite at the wall ($x=0$).
In particular, corresponding to eq. (\ref{lowt}), it satisfies 
\begin{eqnarray}
\rho_{{\tt mixed}}^{(N)}=
-\rho_{{\tt thermal}}^{(N)},
\quad&\xi=1/4,&\quad x=0.
\label{mixed0}
\end{eqnarray}

The vacuum contribution and the mixed contribution in 
$P_{\perp}^{(N)}$ vanish identically, resulting that the pressure on the wall is the only component in eq. (\ref{emt}) which is uniform, i.e., 
\begin{equation}
P_{\perp}^{(N)}=
P_{\perp\ {\tt thermal}}^{(N)},
\label{pressurewall}
\end{equation}
where the blackbody pressure in eq. (\ref{pressurewall}) 
is given as in eq. (\ref{rpressure}).
Finally,
\begin{equation}
P_{\parallel}^{(N)}=
P_{\parallel\ {\tt vacuum}}^{(N)}+
P_{\parallel\ {\tt mixed}}^{(N)}+
P_{\parallel\ {\tt thermal}}^{(N)},
\label{pressurepara}
\end{equation}
where the thermal and vacuum contributions are given by
eqs. (\ref{rpressure}) and (\ref{veqstate}). The mixed 
contribution is given by (see also eq. (\ref{aexpression2}))
\begin{eqnarray}
&&
P_{\parallel\ {\tt mixed}}^{(N)}=
-2\left(\frac{MT}{2\pi}\right)^{N/2}
\sum_{n=1}^{\infty}\left[(2Tx)^2+n^2\right]^{-(N+4)/4}
\nonumber
\\
&&
\hspace{1.0cm}
\times
\left\{4
\left[(N-1)(\xi-\xi_N)\left(4T^2x^2+n^2\right)-
(\xi-1/4) Nn^2\right]
K_{\frac{N}{2}}
\left(\frac{M}{T}\sqrt{(2Tx)^2+n^2}\right)\right. 
\nonumber
\\
&&
\hspace{1.5cm}
\left. +
16MTx^2(\xi-1/4)
\sqrt{(2Tx)^2+n^2}
K_{\frac{N-2}{2}}
\left(\frac{M}{T}\sqrt{(2Tx)^2+n^2}
\right)\right\}.
\label{mixed22}
\end{eqnarray}
Corresponding to eqs. (\ref{lowt}) and (\ref{mixed0}),
\begin{eqnarray}
P_{\parallel\ {\tt mixed}}^{(N)}=
-P_{\parallel\ {\tt thermal}}^{(N)},
\quad&\xi=1/4,&\quad x=0.
\label{mixed000}
\end{eqnarray}

One can check the consistency of the formulas just obtained for the components
of $\left<T^\mu{}^\nu\right>$ by calculating its trace  $\left<T^\mu{}_\mu\right>$. Since the background is flat
the stress-energy-momentum tensor is expected to be traceless 
when $\xi=\xi_{N}$ and $M=0$. Indeed, a straightforward calculation yields  
\begin{equation}
\left<T^\mu{}_\mu\right>^{(N)}_{{\tt thermal}}=
M^2\left<\phi^2\right>^{(N)}_{{\tt thermal}},
\label{bbtrace}
\end{equation}
and
\begin{equation}
\left<T^\mu{}_\mu\right>^{(N)}_{{\tt vacuum}}=
M^2\left<\phi^2\right>^{(N)}_{{\tt vacuum}},
\hspace{0.9cm}
\left<T^\mu{}_\mu\right>^{(N)}_{{\tt mixed}}=
M^2\left<\phi^2\right>^{(N)}_{{\tt mixed}},
\hspace{0.9cm} \xi=\xi_{N}.
\label{vmtrace}
\end{equation}
Using now eq. (\ref{prescription}), one gets
\begin{equation}
\left<T^\mu{}_\mu\right>^{(N)}=
M^2\left<\phi^2\right>^{(N)},
\hspace{0.9cm} \xi=\xi_{N},
\label{trace}
\end{equation}
with the mean square field fluctuation given in eq. (\ref{phi2}). 

The asymptotic behaviour at low temperature (or near the wall) of $\rho^{(N)}$ and $P_{\parallel}^{(N)}$ 
in eq. (\ref{emt}) can be
obtained by simply setting $Tx=0$ in the expressions for their 
mixed contributions. After some algebra it results,
\begin{eqnarray}
P_{\parallel}^{(N)}=-\rho^{(N)}=
-\rho^{(N)}_{{\tt vacuum}}
-(1-4\xi)P_{\perp\ {\tt thermal}}^{(N)},
&\qquad &
Tx\ll 1,
\label{lowtemt}
\end{eqnarray}
up to zeroth order in $Tx$. It should be pointed out that,
unlike eq. (\ref{lowtphi2}), the correction to
the vacuum contribution in eq. (\ref{lowtemt})
does not depend on $x$. This can be understood by recalling that
$\left<T^\mu{}^\nu\right>$ is obtained from a propagator
by applying to the latter a differential operator (see eq. \ref{prescription}). 
Regarding $P_{\perp}^{(N)}$, 
eq. (\ref{pressurewall}) holds everywhere and at all temperatures.

The asymptotic behaviour of the energy density $\rho^{(N)}$ in eq. (\ref{emt}) at high temperature (or deep in the bulk) can be found
by proceeding as in deriving eqs. (\ref{hol1}) and (\ref{classvf}).
Namely, noting eq. (\ref{energy}), the sums in the expression for
the mixed contribution in eq. (\ref{aexpression1}) are replaced by
integrations that can be solved \cite{gra07}. Up to exponentially 
small corrections one finds,
\begin{eqnarray}
\rho^{(N)}=
\rho^{(N)}_{{\tt thermal}}+
\rho^{(N)}_{{\tt class}},&\qquad &
Tx\gg 1,
\label{hol2}
\end{eqnarray}
where
\begin{equation}
\rho^{(N)}_{{\tt class}}=
\left(1-4\xi\right)
\left[2\pi(N-2)
\langle\phi ^{2}\rangle^{(N+1)}_{{\tt vacuum}}
+M^{2}\langle\phi ^{2}\rangle^{(N-1)}_{{\tt vacuum}}
\right]T,
\label{classed0}
\end{equation}
holding also for $N=2$.
Observing now the expression for the vacuum energy density
in eq. (\ref{venergy}), it follows from eq. (\ref{classed0}) when $N>2$,
\begin{equation}
\left(\xi-\xi_{N-1}\right)\rho^{(N)}_{{\tt class}}=
\left(\xi-1/4\right)
\left[\rho^{(N-1)}_{{\tt vacuum}}-
\frac{M^2c^2}{(N-2)\hbar^2}\langle\phi ^{2}\rangle^{(N-1)}_{{\tt vacuum}}
\right]\frac{T}{\hbar c}.
\label{classed}
\end{equation}
Note that $\xi_{N-1}$ in eq. (\ref{classed}) is the conformal coupling
in $N-1$ spacetime dimensions (cf. eq. (\ref{cc})), and that 
dimensionful  $\hbar$ and $c$ were reintroduced (as in eq. (\ref{classvf})) to 
study the ``classical''
nature of $\rho^{(N)}_{{\tt class}}$ shortly. 
The statement in eq. (\ref{classed}) is another instance in the paper of a local dimensional reduction, i.e.,  corrections to  
$N$-dimensional blackbody expressions are obtained from 
($N-1$)-dimensional
vacuum expectation values of local quantities.

Turning now to the nature of $\rho^{(N)}_{{\tt class}}$ in eq. (\ref{classed}),
by considering eq. (\ref{venergy}) and that when $M=0$ 
the vacuum fluctuation $\langle\phi ^{2}\rangle^{(N)}_{{\tt vacuum}}$ depends on $\hbar$ and $c$ only
through the overall factor $\hbar c$ (see discussion in the paragraph containing eq. (\ref{4classvf})), it follows that $\rho^{(N)}_{{\tt class}}$ will be classical only if the field is massless: by setting $M\rightarrow 0$
in eq. (\ref{classed}), the
overall factor $\hbar c$ that arises in  $\rho^{(N-1)}_{{\tt vacuum}}$ 
will cancel that in $T/\hbar c$.

The same line of reasoning applied to eqs. (\ref{pressurepara}) and (\ref{aexpression2}) yields, up to exponentially small corrections,
\begin{equation}
P_{\parallel}^{(N)}=
P_{\parallel\ {\tt thermal}}^{(N)}+
P_{\parallel\ {\tt class}}^{(N)},
\qquad Tx\gg 1,
\label{apressure}
\end{equation}
where,
\begin{equation}
P_{\parallel\ {\tt class}}^{(N)}=
P_{\parallel\ {\tt vacuum}}^{(N-1)}
\ \frac{T}{\hbar c}
\label{cpressure}
\end{equation}
(see eqs. (\ref{rpressure}) and 
(\ref{veqstate})).
Again, local dimensional reduction operates and 
$P_{\parallel\ {\tt class}}^{(N)}$ is classical only if
the field is massless. 

Noticing  eqs. (\ref{classvf}), (\ref{veqstate}), (\ref{classed}) and (\ref{cpressure}),
it follows the equation of state,
\begin{equation}
\left(\xi_{N-1}-\xi\right)\rho^{(N)}_{{\tt class}}=
\left(\xi-1/4\right)
\left[P_{\parallel\ {\tt class}}^{(N)}+
\frac{M^2}{N-2}\langle\phi ^{2}\rangle^{(N)}_{{\tt class}}
\right],
\label{classes}
\end{equation}
relating corrections to the blackbody contributions. If one is interested in $N=2$,
eq. (\ref{classes}) must be multiplied by $N-2$ before setting $N=2$
(cf. eq. (\ref{classed0})). 

Before ending this section it should be remarked that,
because of eq. (\ref{pressurewall}),  
\begin{equation}
P_{\perp\ {\tt class}}^{(N)}=0,
\label{cpressurewall}
\end{equation}
which is consistent with dimensional reduction since
$P_{\perp\ {\tt vacuum}}^{(N)}=0$ for arbitrary $N$. If the wall parallel to that at $x=0$ were brought from infinity to a finite distance, the zero in 
eq. (\ref{cpressurewall}) should give place to the 
well known thermal Casimir pressure \cite{bro69,mil01,sus11}.

\section{Ultralight scalar radiation}
\label{high temperature}

When $T\gg M$, the leading order expressions for 
$\left<\phi^2\right>^{(N)}$ and $\left<T^\mu{}^\nu\right>^{(N)}$ in eqs. (\ref{phi2})
and (\ref{emt}) can be obtained simply by taking 
$M\rightarrow 0$ and considering eq. (\ref{sarguments}).
For completeness, the relevant quantities are given below. They are the masless
expressions,
\begin{eqnarray}
\langle\phi^{2}\rangle_{{\tt vacuum}}^{(N)}&=&-\frac{1}{(2\pi^{1/2})^N}\Gamma\left(\frac{N-2}{2}\right)x^{2-N},
\nonumber
\\
\langle\phi ^{2}\rangle_{{\tt mixed}}^{(N)}&=&-\frac{1}{2\pi^{N/2}}\Gamma\left(\frac{N-2}{2}\right)T^{N-2}
\sum_{n=1}^{\infty}\left[(2Tx)^2+n^2\right]^{(2-N)/2},
\nonumber
\\
\langle\phi ^{2}\rangle_{{\tt thermal}}^{(N)}&=&\frac{1}{2\pi^{N/2}}\Gamma\left(\frac{N-2}{2}\right)\zeta(N-2)T^{N-2},
\label{phis3}
\end{eqnarray}
and
\begin{eqnarray}
\rho^{(N)}_{{\tt mixed}}
=\frac{4}{\pi^{N/2}}\Gamma\left(\frac{N}{2}\right)T^{N}
\sum_{n=1}^{\infty}\left[(2Tx)^2+n^2\right]^{-(N+2)/2}\hspace{2cm}&&
\nonumber
\\
\times\left[(N-1)(\xi-\xi_N)\left(4T^2x^2+n^2\right)-\xi Nn^2\right],&&
\nonumber
\\
P_{\parallel\ {\tt mixed}}^{(N)}
=-\frac{4}{\pi^{N/2}}\Gamma\left(\frac{N}{2}\right)T^{N}
\sum_{n=1}^{\infty}\left[(2Tx)^2+n^2\right]^{-(N+2)/2}\hspace{2cm}&&
\nonumber
\\
\times\left[(N-1)(\xi-\xi_N)\left(4T^2x^2+n^2\right)-(\xi-1/4) Nn^2\right].&&
\label{mixed11}
\end{eqnarray}
Any component of massless $\left<T^\mu{}^\nu\right>^{(N)}$
(cf. ref. \cite{lor14})
can be obtained by using eqs. (\ref{phis3}) and (\ref{mixed11}) in the general expressions found in the previous section. 
It is worth remarking that,
noticing $\left<\phi^2\right>^{(N)}$ given by eq. (\ref{phi2}) and the
functional dependence on $N$ in eqs. (\ref{phis3}), one must have
$N>3$ in order to avoid divergences in the mean square field fluctuation
of hot massless scalar radiation (if $T=0$, one must have $N>2$).

\section{Neumann's boundary condition}
\label{neumann}

Another boundary condition  also used to prevent fluxes 
through the plane wall is Neumann's. Instead of taking $\phi$ itself to
vanish at the wall, one takes,
\begin{equation}
\frac{\partial \phi}{\partial x}=0,
\qquad x=0.
\label{nbc}
\end{equation}
Accordingly, the new Feynman propagator $G_{{\cal F}}({\rm x},{\rm x}')$ at finite temperature $T$ is obtained by solving again eq. (\ref{fe0}), but now
observing eq. (\ref{nbc}) (see appendix \ref{ftpropagator}).
The corresponding expression for $G_{{\cal F}}({\rm x},{\rm x}')$ can be
obtained from that in eq. (\ref{pro}) by replacing the minus sign between
the terms containing the modified Bessel functions by a plus sign.
It follows then that eq. (\ref{fp}) holds with 
$G_{0}({\rm x},{\rm x}')$ and $G_{{\tt thermal}}({\rm x},{\rm x}')$
still denoting the vacuum and the thermal propagators
in Minkowski spacetime, respectively, whereas  
$G_{{\tt vacuum}}({\rm x},{\rm x}')$ and 
$G_{{\tt mixed}}({\rm x},{\rm x}')$ are the negative of the 
Dirichlet ones. Noting these modifications the renormalized propagator
is given as in eq. (\ref{rfp}). Consequently the ensemble average 
of a quantity $\cal{A}$ (cf. eq. (\ref{bm})) near the Neumann wall
can be obtained as follows,
\begin{equation}
\left<\cal{A_{N}}\right>=
-\left<\cal{A_{D}}\right>_{{\tt vacuum}}
-
\left<\cal{A_{D}}\right>_{{\tt mixed}}
+
\left<\cal{A}\right>_{{\tt thermal}},
\label{nbm}
\end{equation} 
where 
$\left<\cal{A_{D}}\right>_{{\tt vacuum}}$ and 
$\left<\cal{A_{D}}\right>_{{\tt mixed}}$
are expressions for Dirichlet's wall. It is worth remarking
that eq. (\ref{nbm}) generalizes to finite temperature $T$ the 
well known identity, 
\begin{eqnarray}
\left<\cal{A_{N}}\right>_{{\tt vacuum}}=
-\left<\cal{A_{D}}\right>_{{\tt vacuum}}.
\label{ndvacuum}
\end{eqnarray}
As $T\rightarrow 0$, eq. (\ref{nbm}) leads to eq. (\ref{ndvacuum}).
Clearly 
\begin{equation}
\left<\cal{A_{N}}\right>_{{\tt mixed}}=
-\left<\cal{A_{D}}\right>_{{\tt mixed}}
\label{ndmixed}
\end{equation}
also holds.

Considering eq. (\ref{nbm}) and the corresponding Dirichlet expressions
computed in the previous sections, one can determine the 
asymptotic thermal behaviours
of 
$\left<\phi^2\right>$ and $\left<T^\mu{}^\nu\right>$
for the Neumann boundary condition at the plane wall. In order to simplify the notation
and to allow ready comparison with the Dirichlet boundary condition, in the rest
of the section expressions denoted as $\left<\cal{A}\right>$,
$\left<\cal{A}\right>_{{\tt vacuum}}$ and 
$\left<\cal{A}\right>_{{\tt mixed}}$ correspond to those for Neumann's boundary condition (cf. eqs. (\ref{ndvacuum}) and (\ref{ndmixed})).

Begining with the mean square field fluctuation, now 
$\langle\phi ^{2}\rangle_{{\tt mixed}}^{(N)}=
\langle\phi ^{2}\rangle_{{\tt thermal}}^{(N)}$ at $x=0$, 
replacing eq. (\ref{lowt}) and leading to 
\begin{eqnarray}
\langle\phi ^{2}\rangle^{(N)}=
\langle\phi^{2}\rangle_{{\tt vacuum}}^{(N)}+
2\langle\phi ^{2}\rangle_{{\tt thermal}}^{(N)},
&\qquad &
Tx\ll 1.
\label{nlowtphi2}
\end{eqnarray}
Thus eq. (\ref{nlowtphi2}) shows that, unlike the Dirichlet $\langle\phi ^{2}\rangle$  (cf. eq. (\ref{lowtphi2})), the thermal behaviour of 
$\langle\phi ^{2}\rangle$ near the Neumann wall is (in leading order) uniform, namely, twice that of a blackbody. Perhaps it should be noted
that Neumann vacuum fluctuations are plagued by the same boundary divergences as their Dirichlet counterparts (more precisely, they differ by a minus sign).

When $Tx\gg 1$
(high temperature or deep in the bulk), because of eqs. (\ref{ndvacuum}) and (\ref{ndmixed}) one can see that the local version of dimensional reduction also operates in the bulk of a ``Neumann cavity''. Regarding 
$\langle\phi ^{2}\rangle$, eqs. (\ref{hol1}) and  (\ref{classvf}) 
still apply, with the vacuum expectation value in eq. (\ref{classvf}) corresponding now to Neumann's boundary condition (cf. eq.  (\ref{ndvacuum})).

The ensemble average of the stress-energy-momentum tensor is given again
by the sum of diagonal terms as in eq. (\ref{prescription}), resulting in
eq. (\ref{emt}) now adapted for Neumann's boundary condition at the plane wall (see eqs.
(\ref{nbm}), (\ref{ndvacuum}) and (\ref{ndmixed})).
Clearly eqs. (\ref{venergy}), (\ref{veqstate}) and (\ref{pressurewall})
remain unchanged, whereas eqs. (\ref{mixed0}) and (\ref{mixed000}) are replaced by (for $\xi=1/4$ and $x=0$)
$\rho_{{\tt mixed}}^{(N)}=\rho_{{\tt thermal}}^{(N)}$ and
$P_{\parallel\ {\tt mixed}}^{(N)}=P_{\parallel\ {\tt thermal}}^{(N)}$.
Regarding traces, one sees that eqs. (\ref{vmtrace}) and (\ref{trace})
still hold.

At low temperature (or near the Neumann wall), 
$\rho^{(N)}$ and $P_{\parallel}^{(N)}$ are given by,
\begin{eqnarray}
\rho^{(N)}=
\rho^{(N)}_{{\tt vacuum}}
+(2N-3+4\xi)P_{\perp\ {\tt thermal}}^{(N)}+2M^2\langle\phi ^{2}\rangle_{{\tt thermal}}^{(N)},
&\qquad &
Tx\ll 1,
\label{nlowtemt}
\end{eqnarray}
and
\begin{eqnarray}
P_{\parallel}^{(N)}=
-\rho^{(N)}_{{\tt vacuum}}
+(3-4\xi)P_{\perp\ {\tt thermal}}^{(N)},
&\qquad &
Tx\ll 1,
\nonumber
\end{eqnarray}
up to zeroth order in $Tx$. These equations replace eq. (\ref{lowtemt}).
It should be noted in
particular that the first equality in eq. (\ref{lowtemt}) no longer holds near the Newmann wall. As mentioned earlier in this section
``local dimensional reduction'' also operates for the Neumann boundary
condition, i.e., eqs. (\ref{hol2}) and (\ref{classed}) hold with the
vacuum expectation values corresponding to Neumann's vacuum (cf. eqs. (\ref{ndvacuum})). Perhaps this is a good place to remark that
by setting $N=4$ and $M=0$, the 
expressions in this section reproduce consistently the results in refs.
\cite{ken80,tad86}.

It is worth mentioning that a curiosity spotted in ref. \cite{ken80}
is associated to the fact that the ``local dimensional reduction''
operates for both boundary conditions at the plane wall. Quoted from
ref. \cite{ken80}: ``... the correction to the Planckian energy density at high temperature is for minimal coupling, ... , exactly three times that for conformal coupling, irrespective of the boundary conditions ... ''. Noting that eq. (\ref{classed0}) applies equally to both boundary conditions, it follows that,
\begin{equation}
\rho^{(N)}_{{\tt class}}(\xi=0)=
\left(N-1\right)\ \rho^{(N)}_{{\tt class}}(\xi=\xi_{N}).
\label{curiosity}
\end{equation}
Now one sees that the curiosity spotted in ref. \cite{ken80} is
the content of eq. (\ref{curiosity}) for $N=4$.

Comparison of 
$d\rho^{(N)}/dT$ for the Dirichlet and Neumann boundary conditions
(denoted below by $c_{\cal D}$ and $c_{\cal N}$, respectively)
reveals contrasting features. For simplicity a massless scalar field
will be considered (i.e., $M=0$).
Noting eqs. (\ref{lowtemt}), (\ref{nlowtemt}) and (\ref{eqstate}), it follows that,
\begin{eqnarray}
&&c_{\cal D}=\frac{1-4\xi}{N-1}c_{V},
\qquad
c_{\cal N}=\frac{2N-3+4\xi}{N-1}c_{V},
\qquad
Tx\ll 1,
\label{sheat}
\end{eqnarray}
where $c_{V}:=d\rho^{(N)}_{{\tt thermal}}/dT$
is the familiar (nonnegative) blackbody specific heat per unity of volume. By examining eq. (\ref{sheat}) one sees that 
when $\xi>1/4$, $c_{\cal D}<0$ and $c_{\cal N}>0$.
The physical interpretation of these inequalities is as follows. 
When $\xi>1/4$, for increasing temperature $T$, the energy density
near the Dirichlet wall decreases, whereas near the Neumann wall it 
increases. Deep in the bulk, both $c_{\cal D}$ and $c_{\cal N}$
behave in leading order as $c_{V}$ (cf. eq. (\ref{hol2})), i.e.,
the energy density increases as the temperature increases.

\section{Conclusion}
\label{conclusion}

Blackbody radiation of a massive scalar field is a topic of interest both in cosmology and high energy physics.
This work examined the local behaviour (in contrast to the usual global approach) of hot scalar radiation  near infinite plane Dirichlet and
Neumann walls in flat spacetime with $N$ dimensions. The Dirichlet  
boundary condition (which is conformal invariant) and the 
Neumann boundary condition (which is not conformal invariant) aimed to model 
the  real reflecting  wall of a large cavity. 
The scalar field $\phi$ was taken to be neutral, with mass $M$, and arbitrarily coupled to the absent curvature (arbitrary $\xi$).
Using the ``point-splitting'' procedure 
and the Schwinger ``proper time'' representation of the Feynman propagator
at finite temperature $T$, 
new formulas for 
$\left<\phi^2\right>$ and $\left<T^\mu{}^\nu\right>$ were derived,
and those for $\left<T^\mu{}^\nu\right>$ were shown to 
reproduce known results when $N=4$, $M=0$, and $\xi=0$ or $1/6$.

The ensemble averages 
$\left<\phi^2\right>$ and $\left<T^\mu{}^\nu\right>$
were expressed as in eq. (\ref{bm}) where the vacuum contributions
revealed divergences at the wall that are typical of  
the use of idealized boundary conditions. It was mentioned 
that divergences of these kind do not affect the other contributions in eq. (\ref{bm}), both depending on $T$ and vanishing when $T=0$.
It is worth recalling that if the distribution in 
eq. (\ref{n}) were used to compute averages, only the thermal contribution (the blackbody contribution) would appear in eq. (\ref{bm}).
The term denoted by ``mixed'' in eq. (\ref{bm}) has 
hybrid nature, and connects the vacuum and the blackbody contributions.
Such a hybrid nature 
is the ingredient responsible for contrasts in the thermal behaviour corresponding to the Dirichlet and Neumann
boundary conditions.

The asymptotic behaviours of $\left<\phi^2\right>$ and $\left<T^\mu{}^\nu\right>$ at low ($Tx\ll 1$) and high ($Tx\gg 1$) temperature were thoroughly studied. 
At low temperature, the dependence on $T$ was shown to be given by the leading correction to the Dirichlet or Neumann vacuum contributions.
Regarding the regime of high temperature, it was  shown
that corrections 
to the blackbody contributions 
carry Planck's constant if $M\neq 0$ (i.e., they are not ``classical'', in general). Another feature spotted is a ``local dimensional reduction'' by
means of which one can say the corrections to blackbody contributions 
by looking at vacuum contributions 
in one less dimension.
At this point it should be stressed that the ``local dimensional reduction''
is present for massless and massive fields, and for both boundary conditions
considered, suggesting that it may be a common feature of more general setups.

It is worth remarking that although nowhere in the text any restriction on the values of the curvature coupling parameter 
$\xi$ was imposed, under certain plausible assumptions thermodynamic arguments in ref. \cite{lor14} suggest that not all values of $\xi$ are consistent with stable thermodynamic equilibrium.


An extension of this work would be to look at the generality of the ``local dimensional reduction'' when the background
is not strictly flat (especially when an event horizon is present).
Another pertinent extension would be to study
how the ``local dimensional reduction'' operates when
the plane wall is replaced by a spherical shell.
The study of a charged scalar, as well as of fields of higher spins, may also reveal new interesting effects.

\appendix
\section{Finite temperature propagator}
\label{ftpropagator}
The geometry of the 
$N$-dimensional spacetime is given by
\begin{equation}
ds^2=dt^2-dx^2-dy^2-dz^2-\cdots.
\label{metric1}
\end{equation}
The coordinate $x_{1}:=it$ is taken to be real with period 
$\beta:=1/T$. By convenience $x_{0}:=ix$ is also analytically continued  to real values. Considering further $x_{2}:=y$ and $x_{3}:=z$, eq. (\ref{metric1}) becomes,
\begin{equation}
ds^2=dx_{0}^2-dx_{1}^2-dx_{2}^2-dx_{3}^2-\cdots-dx_{N-1}^2,
\label{metric2}
\end{equation}
which with the boundary condition,
\begin{equation}
\psi(x_0,x_1,x_2,\cdots,x_{N-1})
=\psi(x_0,x_1+\beta,x_2,\cdots,x_{N-1}),
\label{bc1}
\end{equation}
characterizes a cylindrical spacetime. The Dirichlet boundary
condition at $x=0$ is implemented by taking,
\begin{equation}
\psi(x_0=0,x_1,x_2,\cdots,x_{N-1})=0.
\label{bc2}
\end{equation}
Since
$\Box_{{\rm x}}:=\partial^2_0-\partial^2_1-\partial^2_2-\partial^2_3-
\cdots-\partial^2_{N-1}$,
the eigenfunctions of the operator
$\Box_{{\rm x}}+M^{2}$ are given by 
\begin{equation}
\psi_{\omega,{\bf k}}({\rm x})=\eta\sin(\omega x_{0})
\exp{(i{\bf k}\cdot {\bf x}}),
\label{efunctions}
\end{equation}
${\bf x}:=(x_1,x_2,\cdots,x_{N-1})$, and where $\eta$, $\omega$ as well as the components of ${\bf k}:=(k_1,k_2,\cdots,k_{N-1})$ are constants. The corresponding eigenvalues are
\begin{equation}
E_{\omega,{\bf k}}={\bf k}\cdot{\bf k}-\omega^2+M^2,
\label{evalues}
\end{equation}
and due to eq. (\ref{bc1}) $k_1=2\pi n/\beta$ with $n$ an integer.
The sine function in eq. (\ref{efunctions}) ensures that 
eq. (\ref{bc2}) holds.

One can  check now that the Feynman propagator is given by
\begin{equation}
G_{{\cal F}}({\rm x},{\rm x}')=
-i\sum_{n=-\infty}^{\infty}\int_{0}^{\infty}d\tau
\int_{0}^{\infty}d\omega
\int_{-\infty}^{\infty}dk_{2}\ \cdots
\int_{-\infty}^{\infty}dk_{N-1}
e^{-i\tau E_{\omega,{\bf k}}}\psi_{\omega,{\bf k}}({\rm x})
\psi^{*}_{\omega,{\bf k}}({\rm x}'),
\label{feynman}
\end{equation}
simply by applying $\Box_{{\rm x}}+M^{2}$ to eq. (\ref{feynman}),
\begin{eqnarray}
&&\left(\Box_{{\rm x}}+M^{2}\right)
G_{{\cal F}}({\rm x},{\rm x}')=
\sum_{n=-\infty}^{\infty}
\int_{0}^{\infty}d\omega
\int_{-\infty}^{\infty}dk_{2}\ \cdots
\int_{-\infty}^{\infty}dk_{N-1}
\psi_{\omega,{\bf k}}({\rm x})
\psi^{*}_{\omega,{\bf k}}({\rm x}')
\nonumber
\\
&&
\hspace{4.0cm}\times\int_{0}^{\infty}d\tau\frac{d}{d\tau}
e^{-i\tau E_{\omega,{\bf k}}}.
\label{propertime}
\end{eqnarray}
To make the integration over the ``proper time'' $\tau$
in eq. (\ref{propertime}) convergent, $M^2$ in Eq. (\ref{evalues}) is taken to have an infinitesimal negative imaginary part, resulting that the integration over $\tau$ yields minus unity. By considering eq. (\ref{efunctions})
in eq. (\ref{propertime}) with
$\eta$ such that 
\begin{equation}
|\eta|^2=2^{3-N}\pi^{1-N}\beta^{-1},
\label{normalization}
\end{equation}
the integrations over the components of ${\bf k}$
give the usual representation of the $\delta$-function.
Noting that ${\rm x}$ and ${\rm x}'$ are at the same side of the
wall, using Poisson's formula,
\begin{equation}
\sum_{n=-\infty}^{\infty}\delta(\lambda-2\pi n)=
\frac{1}{2\pi}\sum_{n=-\infty}^{\infty} e^{-in\lambda},
\label{poisson}
\end{equation}
and that \cite{arf85} 
$$\frac{2}{\pi}\int_{0}^{\infty}d\omega\sin(\omega x)\sin(\omega x')=
\delta(x-x')-\delta(x+x'),$$
one sees that the right hand side of eq. (\ref{propertime}) is indeed minus the delta function in the $N$-dimensional cylindrical spacetime 
(cf. eqs. (\ref{metric2}) and (\ref{bc1})).

In order to arrive in eq. (\ref{pro}), eq. (\ref{efunctions}) is used in eq. (\ref{feynman}) noticing  eq. (\ref{normalization}). In so doing a factor
arises that can be conveniently manipulated as follows,
\begin{eqnarray}
\sum_{n=-\infty}^{\infty}
e^{-i\tau (4\pi^2 n^2/\beta^2)+i(2\pi n/\beta)(x_{1}-x'_{1})}&=&
\sum_{n=-\infty}^{\infty}\int_{-\infty}^{\infty}d\lambda\
\delta(\lambda-2\pi n)\
e^{-i\tau (\lambda^2/\beta^2)+i(\lambda/\beta)(x_{1}-x'_{1})}
\nonumber
\\
&=&
\frac{1}{2\pi}\sum_{n=-\infty}^{\infty}\int_{-\infty}^{\infty}d\lambda
e^{-i\tau (\lambda^2/\beta^2)+i(\lambda/\beta)(x_{1}-x'_{1}-n\beta)},
\nonumber
\end{eqnarray}
where eq. (\ref{poisson}) has been used in the last step.
Now all integrations other than that over $\tau$ 
can be promptly evaluated, and the final  integration over $\tau$ yields the modified Bessel functions \cite{gra07}.
The expression in
eq. (\ref{pro}) is obtained by analytically continuing back
to reall values of $t=-ix_{1}$ and $x=-ix_{0}$, and after going back to the original coordinates in eq. (\ref{metric1}).

Instead of eq. (\ref{bc2}), 
the Neumann boundary condition in eq. (\ref{nbc}) is implemented by taking
\begin{eqnarray}
&&\frac{\partial\psi}{\partial x_0}=0,
\qquad x_0=0.
\nonumber
\end{eqnarray}
Accordingly, 
the eigenfunctions of the operator
$\Box_{{\rm x}}+M^{2}$ are now given by 
\begin{equation}
\psi_{\omega,{\bf k}}({\rm x})=\eta\cos(\omega x_{0})
\exp{(i{\bf k}\cdot {\bf x}}),
\label{nefunctions}
\end{equation}
replacing eq. (\ref{efunctions}). Considering eq. (\ref{nefunctions}),
one follows the same line of reasoning as above, ending with eq. (\ref{pro})
but now with the minus sign into the square bracket giving place to a plus sign.

\section{Alternative expressions for certain averages}
\label{aexpressions}

The following formulas are alternative expressions to
$\rho^{(N)}_{{\tt mixed}}$ and $P_{\parallel\ {\tt mixed}}^{(N)}$
given in eqs. (\ref{mixed00}) and (\ref{mixed22}),
\begin{eqnarray}
\rho^{(N)}_{{\tt mixed}}=&&
2\left(\frac{MT}{2\pi}\right)^{N/2}
\sum_{n=1}^{\infty}\left[(2Tx)^2+n^2\right]^{-N/4}
\left\{2(1-2\xi)
K_{\frac{N}{2}}
\left(\frac{M}{T}\sqrt{(2Tx)^2+n^2}\right)\right. 
\nonumber
\\
&&
\hspace{-0.8cm}
\left. +
\frac{M}{T}\left[(4Tx)^2(\xi-1/4)-n^2\right]
[(2Tx)^2+n^2]^{-1/2}
K_{\frac{N+2}{2}}
\left(\frac{M}{T}\sqrt{(2Tx)^2+n^2}\right)\right\},
\label{aexpression1}
\end{eqnarray}
and
\begin{eqnarray}
P_{\parallel\ {\tt mixed}}^{(N)}=&&
-2\left(\frac{MT}{2\pi}\right)^{N/2}
\sum_{n=1}^{\infty}\left[(2Tx)^2+n^2\right]^{-N/4}
\left\{2(1-2\xi)
K_{\frac{N}{2}}
\left(\frac{M}{T}\sqrt{(2Tx)^2+n^2}\right)\right. 
\nonumber
\\
&&
\hspace{-0.8cm}
\left. +
16  MTx^2(\xi-1/4)
[(2Tx)^2+n^2]^{-1/2}
K_{\frac{N+2}{2}}
\left(\frac{M}{T}\sqrt{(2Tx)^2+n^2}\right)\right\}.
\label{aexpression2}
\end{eqnarray}
They lead to those in the text 
containing $K_{\nu}(z)$ of lower $\nu$
by considering known identities
relating Bessel functions of different orders \cite{gra07}.

\acknowledgments
This work was partially supported by the
Brazilian research agencies CAPES, CNPq and FAPEMIG. 


\end{document}